\documentclass{article}
\pdfoutput=1
\usepackage{lscape}
\usepackage{enumerate}
\usepackage{amsfonts}
\usepackage{amsmath}
\usepackage{textcomp}
\usepackage{amssymb}
\usepackage{color}
\usepackage{graphicx}
\usepackage{pdflscape}
\usepackage{afterpage}
\usepackage[matrix,arrow,curve]{xy}
\usepackage{array}
\usepackage{indentfirst}

\def\be{\begin{equation}}
\def\ee{\end{equation}}

\makeatletter
\renewcommand*{\@cite@ofmt}{\bfseries\hbox}
\makeatother

\hoffset= - 1.2in         

\begin{document}

\title{\vspace{0.1cm}{\Large {\bf  On measuring the topological charge of anyons}\vspace{.2cm}}
\author{{\bf Andrey Morozov$^{a,b}$\thanks{e-mail: andrey.morozov@itep.ru}}
}
\date{ }
}

\maketitle

\vspace{-5.5cm}

\begin{center}
\hfill IITP/TH-8/24\\
\hfill ITEP/TH-9/24\\
\end{center}

\vspace{3.6cm}

\begin{center}

$^a$ {\small {\it NRC ``Kurchatov Institute'', Moscow 123182, Russia}}\\
$^b$ {\small {\it Institute for Information Transmission Problems, Moscow 127994, Russia}}\\
\end{center}

\vspace{1cm}

\begin{abstract}
In this paper we discuss the principles of measuring topological charge or representation traveling in the set of anyons. We describe the procedure and analyze how it works for the different values of parameters of the theory. We also show how it can be modified to be more effective.
\end{abstract}

\vspace{.5cm}



\section{Introduction}

Topological quantum computer is a perspective model of quantum computers, which should have low probability of errors by construction \cite{Kit,ReviewTQC}. The model of such computer is based on moving quasi-particles called anyons. These particles exist in two-dimensional effective theories and the have different statistics than fermions and bosons. Namely, their interchange can produce nontrivial and, generally speaking, non-abelian operators.

Anyons are described by an effective Chern-Simons theory
\begin{equation}
S_{\text{CS}}=\cfrac{k}{4\pi}\int\limits_{S^3} A\wedge dA+\frac{2}{3}A\wedge A\wedge A.
\end{equation}
This theory is gauge invariant, usually gauge group $SU(N)$ is considered. Anyonic particles in such theory can be transformed under different representation of the gauge group, also called topological charge.

\begin{figure}[h!]
\begin{picture}(150,140)(-200,-55)
\qbezier(-12,0)(-18,-6)(-18,-12)
\qbezier(-10,-2)(-6,-6)(-6,-12)
\qbezier(10,-2)(6,-6)(6,-12)
\qbezier(12,0)(18,-6)(18,-12)
\put(-18,-12){\line(0,-1){34}}
\qbezier(-6,-12)(-6,-18)(-2,-22)
\qbezier(6,-12)(6,-18)(0,-24)
\put(18,-12){\line(0,-1){34}}
\qbezier(0,-24)(-6,-30)(-6,-36)
\qbezier(2,-26)(6,-30)(6,-36)
\put(-6,-36){\line(0,-1){10}}
\put(6,-36){\line(0,-1){10}}
\qbezier(-18,-46)(-18,-52)(-12,-52)
\qbezier(-6,-46)(-6,-52)(-12,-52)
\qbezier(18,-46)(18,-52)(12,-52)
\qbezier(6,-46)(6,-52)(12,-52)
\qbezier(-14,2)(-18,6)(-18,12)
\qbezier(-12,0)(-6,6)(-6,12)
\qbezier(12,0)(6,6)(6,12)
\qbezier(14,2)(18,6)(18,12)
\put(-18,12){\line(0,1){34}}
\qbezier(-6,12)(-6,18)(-2,22)
\qbezier(6,12)(6,18)(0,24)
\put(18,12){\line(0,1){34}}
\qbezier(0,24)(-6,30)(-6,36)
\qbezier(2,26)(6,30)(6,36)
\put(-6,36){\line(0,1){10}}
\put(6,36){\line(0,1){10}}
\qbezier(-18,46)(-18,52)(-12,52)
\qbezier(-6,46)(-6,52)(-12,52)
\qbezier(18,46)(18,52)(12,52)
\qbezier(6,46)(6,52)(12,52)
\put(-30,41){\line(1,0){60}}
\put(-30,61){\line(0,-1){122}}
\put(30,-41){\line(-1,0){60}}
\put(30,-61){\line(0,1){122}}
\put(-30,61){\line(1,0){60}}
\put(-30,-61){\line(1,0){60}}
\put(35,50){\hbox{annihilation/measurement}}
\put(35,0){\hbox{entangling/operations}}
\put(35,-50){\hbox{creation}}
\end{picture}
\caption{Basic principle of quantum algorithm for a topological quantum computer. Pairs of anyons are produced, then they are entangled and then annihilated. Two-bridge link, presented on the picture corresponds to the one-qubit operations.\label{f:anyons}}
\end{figure}
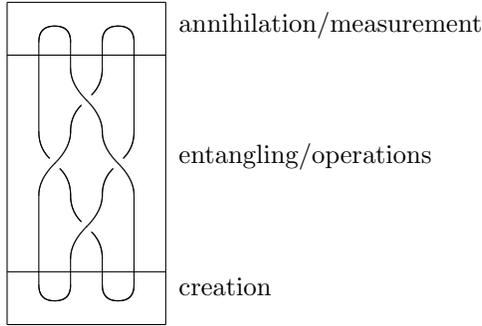

Quantum computations in such a model are made by moving such particles with non-trivial operations made by intertwining their trajectories, see Fig. \ref{f:anyons}. Basic algorithm works as follows. First a pair or several pairs of anyons are produced, then we intertwine their trajectories, finally we try to annihilate pairs of anyons and, depending on the final state of the system, then can either annihilate or not. If they annihilate, then their trajectories form a knot or a link, and, in fact, the probability amplitude of such a process is in fact equal to the knot polynomial of such a knot  \cite{TowTQC,QuantKnots,LargeK}. Intertwining of the trajectories correspond to the quantum $\mathcal{R}$-matrices, which act as elementary operations in topological quantum calculations.

The simplest description of qubit \cite{QuantKnots, LargeK, UniGate} is made from two pairs of anyons, which trajectories form a two-bridge knot. From the point of view of representation theory the whole braid consisting of four trajectories carries trivial representation of the gauge group, since $\mathcal{R}$-matrices do not change the irreducible representations travelling along any number of trajectories (see \cite{HMFLII, Cable, RPaths} for details). However, there can be several trivial representations, appearing in the product of representations corresponding to each anyon. If all anyons correspond to the fundamental representation of the $SU(N)$ group (anyons travelling in the opposite direction then carry anti-fundamental representation), then as a result we get a two-dimensional space of trivial representations. These trivial representations correspond to the different representations travelling in pairs of anyons, rather that the whole braid.

In topological quantum computer there is a special property which does not exist in many other types. Namely, it is possible to measure the state of qubit without changing its state. In other words we can measure the irreducible representation travelling in a braid made of anyon trajectories.  This gives us more freedom in dealing with such a computer.

In this paper we discuss how to measure the irreducible representation or topological charge of anyon and how to make this measurement more precise.

\section{Principle of measurement}

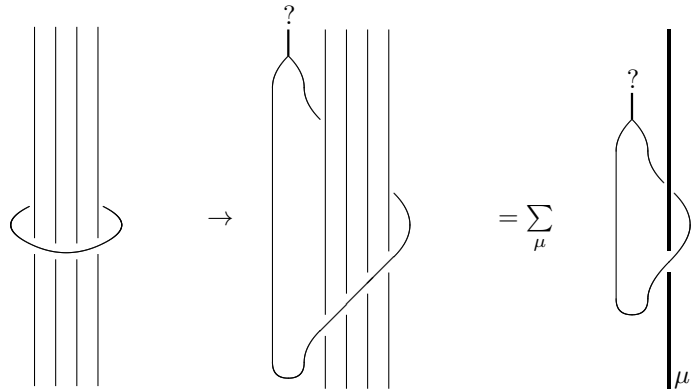
\begin{figure}[h!]
\begin{picture}(250,140)(-200,-55)
\put(-50,0){
\qbezier(14,7)(28,0)(14,-7)
\qbezier(14,-7)(0,-14)(-14,-7)
\qbezier(-14,-7)(-28,0)(-14,7)
\put(-12,-5){\line(0,1){80}}
\put(-4,-7){\line(0,1){82}}
\put(12,-5){\line(0,1){80}}
\put(4,-7){\line(0,1){82}}
\put(-12,-11){\line(0,-1){50}}
\put(-4,-13){\line(0,-1){48}}
\put(12,-11){\line(0,-1){50}}
\put(4,-13){\line(0,-1){48}}
}
\put(0,0){
\hbox{$\rightarrow$}
}
\put(40,0){
\qbezier(-12,-52)(-12,-58)(-6,-58)
\qbezier(0,-52)(0,-58)(-6,-58)
\qbezier(0,-52)(0,-46)(6,-40)
\put(6,-40){\line(1,1){28}}
\qbezier(34,-12)(40,-6)(40,0)
\qbezier(40,0)(40,6)(34,12)
\put(-12,-52){\line(0,1){104}}
\qbezier(-12,52)(-12,58)(-6,64)
\qbezier(0,52)(0,58)(-6,64)
\qbezier(0,52)(0,46)(6,40)
\put(-6,64){\linethickness{0.3mm}\line(0,1){10}}
\put(-8,76){\hbox{?}}
\put(8,-34){\line(0,1){108}}
\put(16,-26){\line(0,1){100}}
\put(24,-18){\line(0,1){92}}
\put(32,-10){\line(0,1){84}}
\put(8,-42){\line(0,-1){20}}
\put(16,-34){\line(0,-1){28}}
\put(24,-26){\line(0,-1){36}}
\put(32,-18){\line(0,-1){44}}
}
\put(110,0){
\hbox{$=\sum\limits_{\mu}$}
}
\put(170,0){
\qbezier(-12,-28)(-12,-34)(-6,-34)
\qbezier(0,-28)(0,-34)(-6,-34)
\qbezier(0,-28)(0,-22)(8,-14)
\put(6,-16){\line(1,1){4}}
\qbezier(8,-14)(16,-6)(16,0)
\qbezier(16,0)(16,6)(10,12)
\put(-12,-28){\line(0,1){56}}
\qbezier(-12,28)(-12,34)(-6,40)
\qbezier(0,28)(0,34)(-6,40)
\qbezier(0,28)(0,22)(6,16)
\put(-6,40){\linethickness{0.3mm}\line(0,1){10}}
\put(-8,52){\hbox{?}}
\put(8,-10){\linethickness{0.4mm}\line(0,1){84}}
\put(8,-18){\linethickness{0.4mm}\line(0,-1){44}}
\put(10,-60){\hbox{$\mu$}}
}
\end{picture}
\caption{Measurement procedure with pair of ancillary anyons. Depending on probability of their annihilation we can find which representaion $\mu$ travels along the braid. \label{f:measimp}}
\end{figure}

To measure the representation travelling in some braid one can take a strand and loop it around the braid. This means that we produce an additional pair of ancillary anyons and intertwine one of them with the studied braid, then we measure the probability of annihilation of this ancillary pair of anyons, see Fig. \ref{f:measimp}. Depending on the representation, travelling in the measured braid, this probability will be different.

As was discussed in \cite{Cable}, instead of a braid, we can consider just one strand but in the higher irreducible representation. Moreover, representation travelling along the braid remains unchanged. Let us consider an example of measurement. We will take $N=2$ and consider a braid made of two anyons in fundamental representation. This means that along the braid can travel either trivial or symmetric representation.

\begin{equation}
[1]\otimes [1]=\emptyset + [2].
\end{equation}

Now let us describe the operators appearing in such braids, using  methods from \cite{Cable, MultiLink, MultiLink1}.

\section{$\mathcal{R}$-matrices}

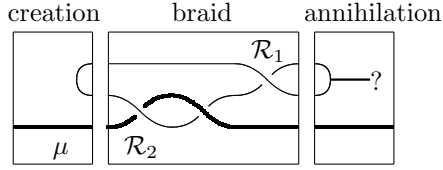
\begin{figure}[h!]
\begin{picture}(100,140)(-255,-55)
\put(0,0){\linethickness{0.4mm}
\qbezier(-24,-12)(-18,-12)(-14,-8)
\qbezier(-10,-4)(-6,0)(0,0)
\qbezier(0,0)(6,0)(12,-6)
\qbezier(12,-6)(18,-12)(24,-12)
\put(24,-12){\line(1,0){24}}
}
\qbezier(-24,0)(-18,0)(-12,-6)
\qbezier(-12,-6)(-6,-12)(0,-12)
\qbezier(0,-12)(6,-12)(10,-8)
\qbezier(14,-4)(18,0)(24,0)
\qbezier(24,0)(30,0)(34,4)
\qbezier(38,8)(42,12)(48,12)
\put(-24,12){\line(1,0){48}}
\qbezier(24,12)(30,12)(36,6)
\qbezier(36,6)(42,0)(48,0)
\put(-18,-22){\hbox{$\mathcal{R}_2$}}
\put(30,14){\hbox{$\mathcal{R}_1$}}
\put(-24,-26){\line(1,0){72}}
\put(-24,-26){\line(0,1){50}}
\put(48,24){\line(-1,0){72}}
\put(48,24){\line(0,-1){50}}
\put(0,28){\hbox{braid}}
\put(-60,-26){\line(1,0){30}}
\put(-60,-26){\line(0,1){50}}
\put(-30,24){\line(-1,0){30}}
\put(-30,24){\line(0,-1){50}}
\put(-62,28){\hbox{creation}}
\qbezier(-30,12)(-36,12)(-36,6)
\qbezier(-36,6)(-36,0)(-30,0)
\put(-30,-12){\linethickness{0.4mm}
\line(-1,0){30}
}
\put(-45,-22){\hbox{$\mu$}}
\put(54,-26){\line(1,0){30}}
\put(54,-26){\line(0,1){50}}
\put(84,24){\line(-1,0){30}}
\put(84,24){\line(0,-1){50}}
\put(50,28){\hbox{annihilation}}
\qbezier(54,12)(60,12)(60,6)
\qbezier(60,6)(60,0)(54,0)
\put(60,6){\linethickness{0.3mm}
\line(1,0){15}
}
\put(75,2){\hbox{?}}
\put(54,-12){\linethickness{0.4mm}
\line(1,0){30}
}
\end{picture}
\caption{Braid we need to study to describe interaction between thick strand carrying studied representation $\mu$ and ancillary anyon. Braid consist of $\mathcal{R}$-matrices of different types -- $\mathcal{R}_1$ corresponds to the crossing between first two strands, and $\mathcal{R}_2$ -- between the second two. \label{f:braid}}
\end{figure}

We want to consider three-strand braid, as in Fig. \ref{f:braid}. This braid consist of two strands in fundamental representation and one strand in either trivial or symmetric representation. The case of trivial representation is very simple. Trivial representation interact trivially with other representations. This means that the state in the first pair of strands remains unchanged, thus the probability of annihilation is equal to $1$.

Now let us discuss the symmetric representation. Since at the beginning we produced a pair of anyons in trivial representation and there is also another strand carrying symmetric representation $[2]$, the whole braid also carries representation $[2]$. However, in such a braid there is a two-dimensional space of such representations. This can be seen, for example, from the following expansion:
\begin{equation}
[1]\otimes[1]\otimes[2]=([2]+\emptyset)\otimes[2]=([4]+[2]+\emptyset)+[2].
\end{equation}
One of symmetric representations comes from trivial representation travelling in the first pair of strands and the other -- to the symmetric representation. The whole braid then correspond to a two by two matrix $B$ which describes how a pair of symmetric representations at the start transform into pair of representations at the end. Since we know that by construction at the start we had trivial representation, element $B[1,1]$ describes probability amplitude of annihilation, and $B[1,2]$ -- of non-annihilation of this ancillary pair of anyons.

\begin{figure}[h!]
\centering
\includegraphics[width=10cm]{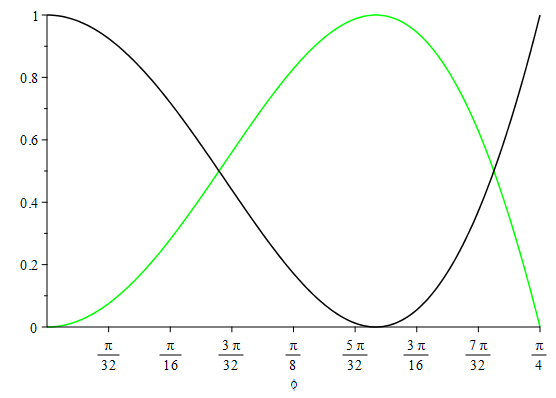}
\caption{Probabilities of pair of ancillary anyons carrying symmetric or trivial representation as functions of $\phi$. Black curve is the probability of measuring trivial representation, green curve -- symmetric representation.
\label{f:graph}}
\end{figure}

To construct this matrix $B$ we need to describe $\mathcal{R}$-matrices, appearing in such a braid. They are described using variable $q$, which is made from the level $k$ of Chern-Simons theory and quantum numbers $[n]_q$, constructed from q:
\begin{equation}
q=e^{\cfrac{2\pi i}{k+2}}, \ \ \ \ [n]_q=\cfrac{q^n-q^{-n}}{q-q^{-1}}.
\end{equation}
It is also important to mention that physical Chern-Simons theories correspond to integer values of $k$. Even further, as was discussed in \cite{LargeK}, $k\geq 6$.
Using techniques described in \cite{Cable, MultiLink, MultiLink1} we can construct matrix $\mathcal{R}^2_2$ appearing in the braid on the left side of Fig. \ref{f:measimp}.
\begin{equation}
\mathcal{R}^2_2=\left(\begin{array}{cc}
\cfrac{q^{2}[6]_q}{[2]_q[3]_q}
&
\sqrt{\cfrac{[4]_q}{[2]_q}}\cfrac{q^4-q^{-2}}{[3]_q}
\\
\sqrt{\cfrac{[4]_q}{[2]_q}}\cfrac{q^4-q^{-2}}{[3]_q}
&
\cfrac{[6]_q}{[2]_q[3]_q}
\end{array}\right).
\label{eq:R22}
\end{equation}
It means that the probability amplitudes of getting trivial or symmetric representations in the product of two strands in the end, or, correspondingly, probability of anyons annihilating or not annihilating, are equal to
\begin{equation}
\begin{array}{l}
A_{\emptyset}=\cfrac{q^{2}[6]_q}{[2]_q[3]_q},
\\
\\
A_{2}=\sqrt{\cfrac{[4]_q}{[2]_q}}\cfrac{q^4-q^{-2}}{[3]_q}.
\end{array}
\end{equation}
We can substitute $q=e^{i\phi}$, then they become
\begin{equation}
\begin{array}{l}
A_{\emptyset}=\cfrac{\sin{6\phi}\sin{\phi}}{\sin{2\phi}\sin{3\phi}}e^{2i\phi},
\\ \\
A_{2}=\sqrt{2\cos{2\phi}}\ 2e^{i\phi}\sin{\phi}=\sqrt{4\cos{2\phi}(1-\cos{2\phi})}e^{i\phi}.
\end{array}
\end{equation}
On the Fig. \ref{f:graph} we present the graphs for the probability of pair of ancillary anyons carrying either representation.

For $k=10$, $\phi=2\pi/(k+2)=\pi/6$, the probability of measuring symmetric representation becomes equal to $1$. It is the perfect parameters of the theory to measure the representation ir topological charge, because in this point ancillary anyons annihilate only if the measured representation was trivial, and do not annihilate if it is a symmetric one. In other cases we can also measure the representation, because since the measurement does not change the state we can make as many measurements as we need, but for $k=10$ this procedure is the simplest.



\section{Measuring with several twists}

\begin{figure}[h!]
\begin{picture}(100,140)(-255,-55)
\put(0,0){
\qbezier(0,-48)(0,-44)(8,-40)
\qbezier(8,-40)(16,-36)(16,-32)
\qbezier(16,-32)(16,-28)(10,-25)
\qbezier(6,-23)(0,-20)(0,-16)
\qbezier(0,-16)(0,-12)(8,-8)
\qbezier(8,-8)(16,-4)(16,0)
\qbezier(16,0)(16,4)(10,7)
\qbezier(6,9)(0,12)(0,16)
\qbezier(0,16)(0,20)(8,24)
\qbezier(8,24)(16,28)(16,32)
\qbezier(16,32)(16,36)(10,39)
\qbezier(0,-48)(0,-44)(8,-40)
\qbezier(8,-40)(16,-36)(16,-32)
\qbezier(16,-32)(16,-28)(10,-25)
\qbezier(6,41)(0,44)(0,48)
\qbezier(-12,-48)(-12,-54)(-6,-54)
\qbezier(0,-48)(0,-54)(-6,-54)
\put(-12,-48){\line(0,1){96}}
\qbezier(-12,48)(-12,54)(-6,60)
\qbezier(0,48)(0,54)(-6,60)
\put(-6,60){\linethickness{0.3mm}\line(0,1){10}}
\put(-8,72){\hbox{?}}
\put(8,-6){\linethickness{0.4mm}\line(0,1){28}}
\put(8,-38){\linethickness{0.4mm}\line(0,1){28}}
\put(8,26){\linethickness{0.4mm}\line(0,1){50}}
\put(8,-42){\linethickness{0.4mm}\line(0,-1){18}}
\put(10,-60){\hbox{$\mu$}}
\put(28,-48){\line(0,1){96}}
\put(28,-48){\line(-1,0){12}}
\put(28,48){\line(-1,0){12}}
\put(32,-5){\hbox{$2b$ twists}}
}
\end{picture}
\caption{Modified measurement procedure with several twists made with one of the ancillary anyons. On the picture there are $2b=6$ twists. \label{f:tws}}
\end{figure}
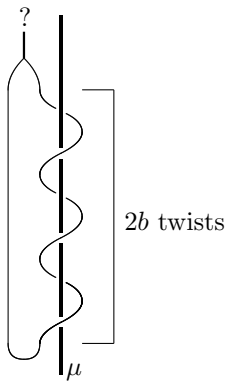

We can modify the measurement algorithm so that it would work better for other values of $k$. For this instead of one twist we can make several twists and try to annihilate ancillary anyons after, like on Fig. \ref{f:tws}. If representation $\mu$, travelling along the measured line is trivial then there is no interaction between it and measuring strands. This means that for any number of twists the probability of annihilation at the end of the loop is still equal to $1$.

If the measured representation is a symmetric one, then $2b$ twists are described by the $b$ power of (\ref{eq:R22}). Matrix element, corresponding to the non-annihilating state is equal to:



\begin{equation}
A_{2}=\sqrt{\cfrac{[4]_q}{[2]_q}}\cfrac{q^b(q^{3b}-q^{-3b})}{[3]_q}
\end{equation}

If we substitute the value of $q=e^{i \phi}$ into this formula we get

\begin{equation}
A_{2}=\sqrt{8\cos{2\phi}}\cfrac{e^{ib\phi}\sin{\phi}\sin{3b\phi}}{\sin{3\phi}}
\label{eq:amp}
\end{equation}

The probability of non-annihilating is maximal as a function of $b$ when

\begin{equation}
3b\phi=\frac{\pi}{2}+\pi n
\label{eq:3bf}
\end{equation}

Since we want both $b$ and $k$ to be integer, there are two series of $k$, satisfying this relation:

\begin{equation}
\left[
\begin{array}{lcl}
\frac{6b\pi}{k+2}=\frac{\pi}{2} & \Rightarrow k=12b-2
\\
\frac{6b\pi}{k+2}=\frac{3\pi}{2} & \Rightarrow k=4b-2
\end{array}
\right.
\end{equation}

 All other series, appearing from (\ref{eq:3bf}), are subseries of these two. Even more than this, the first of the series is also included in the second one. Thus we will consider only this second series $k=4b-2$. Now let us find the probability of ancillary anyons not annihilating for this series. From (\ref{eq:amp}) after substituting $k$ and $\phi=\pi/2b$,

 \begin{equation}
 P_{2}=|A_2|^2=8\cos{\pi/b}\left(\cfrac{\sin{\pi/2b}\sin{3\pi/2}}{\sin{3\pi/2b}}\right)^2=8\cos{\pi/b}\cfrac{1}{\left(\cos{\pi/b}+2\cos^2(\pi/2b))\right)^2}=8\cos{\pi/b}\cfrac{1}{(2cos{\pi/b}+1)^2}
 \end{equation}

 There are two values of $b$, for which this probability is equal to zero, namely $b=1,2$, otherwise for $b=3$, $k=10$, the probability is equal to $1$. After this the derivative of the probability is negative, thus the probability monotonously becomes lower. But the limit of the probability with $b\rightarrow\infty$ is equal to $8/9\thickapprox 0,89$. Therefore,
 for $k=4b-2$, $b\geq 3$, the probability of ancillary anyons not annihilating is higher than $8/9\thickapprox 0,89$.

\section{Conclusion}

In this paper we discussed the possibility of measuring topological charge of several anyons travelling together. The main point is that due to the topological properties of anyons, the topological charge, corresponding to quantum states in topological quantum computer, can be measured without changing the state itself. We found out that the best theory to do this is $SU(2)$ Chern-Simons theory with level $k=10$. We also suggested the modification of measurement algorithm which make it easier to measure the topological charge for the theories with levels $k=4b-2$.

It is also possible to consider more complex configurations for the measurement. The ancillary part can be even further twisted around the studied braid including selfintersections of trajectories of ancillary anyons, which possibly can give even better results. Also it is possible to take ancillary anyons in higher representations, This is probably necessary if we want to distinguish more representations in the studied braid and do it effectively in using small number of measurements. Finally it is interesting to generalize this approach to the $SU(N)$ group. It should be rather straightforward, using methods, described in \cite{Cable, MultiLink, MultiLink1, Tabul, DoubleFat}.

\section*{Aknowledgements}

We are grateful for very useful discussions with S. Mironov and A. Popolitov. This work was supported by the Russian Science Foundation grant No 23-71-10058.



\end{document}